\documentclass[11pt]{article}
\usepackage[margin=2.54cm]{geometry}
\usepackage{algorithmicx}

\usepackage[T1]{fontenc}
\usepackage[utf8]{inputenc}
\usepackage[english]{babel}
\usepackage{amsmath,amssymb, amsthm}
\usepackage[bookmarksopen,colorlinks,linkcolor=blue]{hyperref}
\usepackage{algorithm}
\usepackage{algpseudocode}
\usepackage{authblk}
\usepackage[shortlabels]{enumitem}
\usepackage{graphicx}
\usepackage{enumitem}

\usepackage{lmodern}
\usepackage{color}

\title{New Bounds for the Flock-of-Birds Problem} 

\author{Alexander Kozachinskiy\thanks{Steklov Mathematical Institute, kozlach@mail.ru. A part of this work was done during the Theoretical Foundations of Computer Systems (TFCS 2021) program at the Simons Institute.}
}

\newtheorem{theorem}{Theorem}

\newtheorem{lemma}[theorem]{Lemma}

\newtheorem{definition}{Definition}

\theoremstyle{definition}

\begin{document}
\maketitle

\begin{abstract}
In this paper, we continue a line of work on obtaining succinct population protocols for Presburger-definable predicates. More specifically, we focus on threshold predicates. These are predicates of the form $n\ge d$, where $n$ is a free variable and $d$ is a constant.

For every $d$, we establish a 1-aware population protocol for this predicate with $\log_2 d + \min\{e, z\} + O(1)$ states, where $e$ (resp., $z$) is the number of $1$'s (resp., $0$'s) in the binary representation of $d$ (resp., $d - 1$). This improves upon an upper bound $4\log_2 d + O(1)$ due to Blondin et al. We also show that any 1-aware protocol for our problem must have at least $\log_2(d)$ states. This improves upon a lower bound $\log_3 d$ due to Blondin et al.

\end{abstract}

\section{Introduction}

Population protocols were initially introduced as a model of distributed computation in  large networks of low-memory sensors~\cite{angluin2006computation}. There are also similarities  between population protocols and some models of social networks~\cite{diamadi2001simple} and chemical reactions~\cite{gillespie1977exact}, see a discussion in~\cite{aspnes2009introduction}. Perhaps, population protocols are most known for their deep connection to logic, namely, to Presburger arithmetic. More specifically, there is a theorem that a predicate over the set of natural numbers is definable in Presburger arithmetic if and only if it can be computed by some population protocol~\cite{angluin2007computational}. In this paper, we continue a line of work on the \emph{minimization} of population protocols~\cite{blondin_et_al:LIPIcs:2018:8511,10.1145/3465084.3467912,Blondin2020SuccinctPP}: given a Presburger-definable predicate, what is the minimal size of a population protocol computing it? More specifically, we obtain some new upper and lower bounds for \emph{threshold} predicates.

We start by describing the model of population protocols in more detail.

\medskip

\textbf{The model.} In this paper, we only consider population protocols for unary predicates. On a high level, population protocols are a sort of finite-state distributed algorithms. A population protocol can have an arbitrary natural number $n$ on input. A population protocol \emph{computes} a unary predicate  $R\colon\mathbb{Z}^+\to\{0, 1\}$ if for every $n\in\mathbb{Z}^+$, having $n$ on input, this protocol in some sense ``converges'' to $R(n)$.

In this framework, natural numbers are represented as \emph{populations} of indistinguishable agents (or, in other words, as piles of indistinguishable items). Namely, a natural number $n$ corresponds to a population with $n$ agents. It turns out that this way of representing natural numbers is quite convenient for Presburger arithmetic. Intuitively, this is because to add two numbers in this model, we just have to join the corresponding piles.

A population protocol $\Pi$ is specified by a finite set of states, a transition function mapping pairs of states into pairs of states, and a partition of the set of states into ``0-states'' and ``1-states''. Having a population of $n$ agents on input, $\Pi$ works as follows. First, it puts every agent into an initial state (which is specified in the description of the protocol and is the same for all $n$). Then agents start to \emph{encounter} each other. We assume that the time is discrete and that a single encounter of 2 agents happens in each unit of time. This process is infinite and is not controlled by $\Pi$. However, when two agents meet, their states are updated according to the transition function of $\Pi$ (given a pair of their states before the encounter, it gives a pair of their states after the encounter).

Recall that states of $\Pi$ are partitioned into 1-states and 0-states. This partition is responsible for the opinions of the agents on the value of a predicate. Namely, agents in 1-states (resp., 0-states) ``think'' that $n$ belongs to a predicate (resp., does not belong to a predicate).

  Finally, we clarify what does it mean that $\Pi$ converges to 1 (resp., 0) on $n$. We want all agents to be in 1-states (resp., in 0-states) forever after some finite time. However, it is meaningless to require this for all possible infinite sequences of encounters. For example, it might be that the same two agents meet each other over and over again. There is no chance agents will learn anything about $n$ in this way. So will only consider sequences of encounters that form \emph{fair executions} of our protocol.

  To define this, we first need a notion of a \emph{configuration}.  An $n$-size configuration of $\Pi$ is a function from the set of $n$ agents to the set of states of $\Pi$. In turn, an execution of $\Pi$  is an infinite sequence of configurations such that \emph{\textbf{(a)}} in the first configuration all agents are in the initial state; \emph{\textbf{(b)}} every configuration, except the first one, is obtained from the previous one via some encounter.  In turn, an execution is fair  if for any two configurations $C_1$ and $C_2$ the following holds: if $C_1$ appears infinitely often in our execution, and if $C_2$ is reachable from $C_1$ via some finite sequence of encounters, then $C_2$ also appears infinitely often in our execution.
  Finally, we say that $\Pi$ converges to 1 (resp., 0) on $n$ if all fair executions have this property: all but finitely many configurations of this execution include only agents in 1-states (resp., 0-states).

\bigskip

\textbf{Threshold predicates.} In this paper, we are interested in \emph{threshold predicates}, that is, predicates of the form:
\[R(n) = \begin{cases} 1 & n\ge d,\\ 0 & \mbox{otherwise,}\end{cases}\]
where $d\in\mathbb{Z}^+$. For brevity, below we use the following notation for these predicates: $R(n) = \mathbb{I}\{n \ge d\}$.

The problem of computing this family of predicates by population protocols is called sometimes the \emph{flock-of-birds problem}. This is because of the following analogy due to Angluin et al.~\cite{angluin2006computation}. Imagine a flock of birds, where each bird is equipped with a temperature sensor. Some birds  are sick due to the elevated temperature. Our sensors have very low action radius: two sensors can interact with each other only if they are, say, at most 1 meter apart. Let there be $n$ sick birds. From time to time, two sick birds approach each other sufficiently close so that their sensors can interact. We want to know whether $n$ (the number of sick birds) is at least some threshold $d$. This turns into a problem  of computing the predicate $R(n) = \mathbb{I}\{n \ge d\}$ by a population protocol.

The first population protocol computing this predicate was given in~\cite{angluin2006computation} (in fact, this was the first population protocol ever considered in the literature). It works as follows. Imagine that initially every agent has 1 coin. An agent can hold up to $d - 1$ coins. Consider an arbitrary encounter of two agents. If two agents meet and have less than $d$ coins in total, one of them gets all the coins of the other one. In turn, if they have at least $d$ coins in total, they both become ``converted''. That is, they start to think to $n\ge d$ (initially everybody thinks that $n < d$). Finally, any agent who meets a converted agent also becomes converted.

Let us see why this protocol computes the predicate $R(n) = \mathbb{I}\{n \ge d\}$.
First, assume that $n < d$. Then in the beginning we have less than $d$ coins. The total number of coins is preserved throughout the protocol. In particular, no two agents can have at least $d$ coins in total. Thus, everybody will always think that $n < d$, as required.

Assume now that $n \ge d$. After any sequence of encounters it is still possible to reach a configuration in which everybody is converted. Indeed, go to a configuration with the least possible number of  non-bankrupt agents (bankrupt agents are agents with $0$ coins). In this minimal configuration, any two non-bankrupt agents must have at least $d$ coins in total (otherwise, one could reduce the number of such agents). Thus, it is possible to convert somebody. It remains to pair all the agents one by one with a converted agent.

To finish the argument, consider any fair execution with $n \ge d$ agents. Let $C$ be any configuration which appears infinitely often in this execution. There is a configuration $D$ which is reachable from $C$ and in which everybody is converted. Thus, $D$ must belong to our execution in some place. Starting from this place, everybody will always think that $n\ge d$.

\bigskip

\textbf{1-awareness}. The protocol which we just described has the following feature. If $n < d$, then no agent will ever think that $n\ge d$. In other words, to start thinking that $n\ge d$, an agent must obtain some proof of this fact. In our case, a proof is a physical presence of  $d$ coins.

Blondin et al.~\cite{blondin_et_al:LIPIcs:2018:8511} call this kind of protocols \emph{1-aware protocols}. Formally, they are defined as follows. Let $R\colon\mathbb{Z}^+ \to\{0, 1\}$ be a predicate. We say that a protocol computing this predicate is \emph{1-aware} if the following holds: for every $n$ with $R(n) = 0$, no execution of $n$ agents ever contains an agent thinking that $R(n) = 1$.

It is not hard to see that 1-aware protocols can only compute threshold predicates and the all-zero predicate. Indeed, if it is possible to make one of $n$ agents think that $R(n) = 1$, then the same is possible for all populations with more than $n$ agents. Hence, any predicate $R$ which can be computed by a 1-aware population protocol is monotone: if $R(n) = 1$, then $R(m) = 1$ for every $m > n$.

Thus, 1-aware population protocols are a quite natural model for computing threshold predicates. In this paper, for every $d$ we study the following question: what is the minimal number of states in a 1-aware population protocol, computing the predicate $R(n) = \mathbb{I}\{n\ge d\}$?

\bigskip

\textbf{Our results}. 
Observe first that the protocol of Angluin et al., described above, requires $d + 1$ states. Indeed, in this protocol, agents just memorize how many coins they hold. This is a number from $0$ to $d - 1$. We also need one more state for converted agents.

This can be drastically improved when $d$ is a power of $2$. Consider the same protocol, but forbid  any ``transfers'' of coins unless two agents have the \emph{same number of coins}. Then an agent can hold either $0$ coins or a power of 2 of them. Thus, this modified protocol requires only about $\log_2 d$ states.

When $n\ge d$, it works for the same reasons as before -- minimize the number of non-bankrupt agents and observe that there must 2 of them holding at least $d$ coins (because otherwise they must hold different powers of 2 whose sum is smaller than $d$). In fact, this protocol also works when $d$ is the sum of two powers of 2, but for other $d$ it does not. A problem is that it might be impossible to get two agents with $d$ coins in total (for example, when there are $d = 4 + 2 + 1 = 7$ coins, two agents can hold at most $4 + 2 = 6$ coins). 

Nevertheless, for every $d$, Blondin et al.~\cite{blondin_et_al:LIPIcs:2018:8511} have constructed a 1-aware protocol with $O(\log d)$ states, computing the predicate $R(n) = \mathbb{I}\{n \ge d\}$. Their construction has two steps. First, they solve the problem with a protocol in which encounters can involve not only 2 but up to $\log_2 d$ agents. Second, they show a general result, transforming any protocol with ``crowded''  encounters into a standard protocol. The second part of their argument is rather technical. As a result, they get a protocol with $4\log_2 d + O(1)$ states. Our first result is the following improvement of this upper bound.

\begin{theorem}
\label{thm:upper_bound}
For any $d\in\mathbb{Z}^+$ the following holds: there exists a deterministic 1-aware population protocol with $\log_2 d + \min\{e, z\} + O(1)$ states, computing the predicate $R(n) = \mathbb{I}\{n \ge d\}$. Here $e$ (resp., $z$) is the number of $1$'s (resp., $0$'s) in the binary representation of $d$ (resp., $d - 1$).
\end{theorem}

This upper bound never exceeds $\frac{3}{2}\log_2 d + O(1)$. Indeed, the number of 1's in the binary representation of $d$ is larger at most by one than the number of 1's in the binary representation of $d - 1$. Hence, $e + z$ does not exceed the length of the binary representation of $d - 1$ plus one. This implies that $\min\{e, z\} \le \frac{1}{2}\log_2 d + O(1)$.

In fact,  we devise two different protocols for Theorem \ref{thm:upper_bound}: one with $\log_2 d + e + O(1)$ states, and the other with $\log_2 d + z + O(1)$ states. The first one is given in Section \ref{sec:first_protocol} and the second one is given in Section \ref{sec:second_protocol}. These two protocols require different ideas. Unlike the construction of Blondin et al., our constructions are direct.

Additionally, Blondin et al.~in~\cite{blondin_et_al:LIPIcs:2018:8511} show that any 1-aware protocol computing $R(n) = \mathbb{I}\{n \ge d\}$ must have at least $\log_3 d$ states. Our second result is an improvement of this lower bound.
\begin{theorem}
\label{thm:lower_bound}
For any $d\in\mathbb{Z}^+$ the following holds: any 1-aware population protocol computing the predicate $R(n) = \mathbb{I}\{n \ge d\}$ has at least $\log_2 d + 1$ states.
\end{theorem}
Theorem \ref{thm:lower_bound} is proved in Section \ref{sec:lower_bound}.

\bigskip

\textbf{Other related works}. In this paper we only deal with 1-aware protocols. For general population protocols, the gap between upper and lower bounds is much wider. A simple counting argument shows that for infinitely many $d$, the minimal size of a population protocol computing $R(n) = \mathbb{I}\{n \ge d\}$ is $\Omega(\log^{1/4} d)$. 
We are not aware of any explicit sequence of $d$'s on which this lower bound is attained. Recently, Czerner and Esparza~\cite{10.1145/3465084.3467912} have shown that for \emph{every} $d$, the minimal size of a population protocol computing $R(n) = \mathbb{I}\{n \ge d\}$ is $\Omega(\log\log\log d)$. 

Similar questions have been studied for other predicates. Namely,
Blondin et al.~\cite{Blondin2020SuccinctPP} obtained the following general result. Assume that a predicate $R$ is definable in Presburger arithmetic via some quantifier-free formula of length $l$ (where all constants are written in binary; for example, the predicate $R(n) = \mathbb{I}\{n \ge d\}$ can be given by a formula of length $\log_2 d + O(1)$). Then there is a population protocol with $l^{O(1)}$ states computing $R$.

Let us mention a related line of research which aims to minimize another parameter of population protocols -- \emph{the time of convergence}~\cite{angluin2008fast}.
It is defined as the expected number of encounters until all agents stably have the right opinion on the value of a predicate. We refer the reader to~\cite{alistarh2017time,czerner_et_al:LIPIcs.SAND.2022.11} for the recent results in this area.

\section{Preliminaries}
We only consider population protocols for unary predicates. Moreover, we only define 1-aware population protocols. For more detailed introduction to population protocols, see~\cite{aspnes2009introduction}.

\textbf{Notation}. By $\mathbb{Z}^+$ we denote the set of positive integers. For $n\in\mathbb{Z}^+$, we write $[n] = \{1, 2, \ldots, n\}$. We  also write $A = B\sqcup C$ for three sets $A, B, C$ if $A = B\cup C$ and $B\cap C = \varnothing$. By $2^A$ we mean the power set of a set $A$.

\begin{definition} A \textbf{population protocol} $\Pi$ is a tuple $\langle Q, Q_0, Q_1, q_{init}, \delta\rangle$, where
\begin{itemize}
\item $Q$ is a finite set of states of $\Pi$;
\item $Q_0, Q_1\subseteq Q$ are such that $Q = Q_0\sqcup Q_1$.
\item $q_{init}\in Q$ is the initial state of $\Pi$;
\item $\delta\colon Q^2 \to 2^{Q^2}\setminus\{\varnothing\}$ is the transition function of $\Pi$.
\end{itemize}
We say that $\Pi$ is \textbf{deterministic} if $|\delta(q_1,q_2)| = 1$ for every $q_1, q_2\in Q$.
\end{definition}

Let $\Pi = \langle Q, Q_0, Q_1, q_{init}, \delta\rangle$ be a population protocol. Consider any $n\in\mathbb{Z}^+$. An $n$-size \emph{configuration} of $\Pi$ is a function $C\colon[n]\to Q$. Intuitively, elements of $[n]$ are agents, and the function $C$ maps every agent to the state this agent in. Define the initial $n$-size configuration as $I_n\colon [n]\to Q$, $I_n(i) = q_{init}$ for all $i\in [n]$. A pair of two $n$-size configurations $(C_1, C_2)$ is called a \emph{transition} if there exist $i, j\in [n]$, $i\neq j$ such that
\[(C_2(i), C_2(j)) \in \delta\big((C_1(i), C_1(j))\big) \mbox{ and } C_2(k) = C_1(k) \mbox{ for all } k\in [n]\setminus\{i,j\}.\]
That is, $C_2$ must be obtained from $C_1$ via an encounter of two distinct agents $i$ and $j$. These agents update their states according to $\delta$, and other agents do not change their states.

We stress that 2 agents participating in an encounter are ordered. It is convenient to imagine that one of the agents  ``initiates'' the encounter and the other agent ``responds'' to it.
This is why $\delta$ is defined over ordered pairs of states and not over 2-element subsets of $Q$.

Next, let $C$ and $D$ be two $n$-size configurations. We say that $D$ is \emph{reachable} from $C$ if for some $k\ge 1$ and for some sequence $C_1, C_2, \ldots, C_k$ of configuration we have:
\begin{itemize}
\item $C_1 = C, C_k = D$;
\item for every $1 \le i < k$ we have that $(C_i, C_{i + 1})$ is a transition.
\end{itemize} 

An execution is an infinite sequence $\{C_i\}_{i=1}^\infty$ of configurations such that $C_1 = I_n$ for some $n$ and $(C_i, C_{i + 1})$ is a transition for every $i\in\mathbb{Z}^+$. We call an execution  $E = \{C_i\}_{i=1}^\infty$ \emph{fair} if for every two configurations $C, D$ the following holds: if, first, $C$ occurs infinitely often in $E$, and second, $D$ is reachable from $C$, then $D$ also occurs infinitely often in $E$.

\begin{definition}
\label{def:aware}
Let $R\colon\mathbb{Z}^+\to\{0, 1\}$ be some predicate. 
We say that a population protocol $\Pi = \langle Q, Q_0, Q_1, q_{init}, \delta\rangle$ is a 1-aware population protocol computing $R$ if for any $n\in\mathbb{Z}^+$ the following holds:
\begin{itemize}
\item if $R(n) = 0$, then for every configuration $C$ which is reachable from $I_n$ we have $C([n]) \subseteq Q_0$.
\item if $R(n) = 1$, then for every fair execution $\{C_i\}_{i = 1}^\infty$ which start from  $C_1 = I_n$ there exists $i_0$ such that for every $i\ge i_0$ we have $C_i([n]) \subseteq Q_1$.
\end{itemize}
\end{definition}

\section{Proof of Theorem \ref{thm:lower_bound}}
\label{sec:lower_bound}
Assume that $\Pi = \langle Q, Q_0, Q_1 q_{init}, \delta\rangle$ is a 1-aware population protocol computing the predicate $R(n) = \mathbb{I}\{n\ge d\}$. 
Let $C\colon[n]\to Q$ be a configuration of $\Pi$ and $q\in Q$ be a state. We say that $q$ \emph{can occur} from $C$ if there exists a configuration $D$ of $\Pi$ such that \textbf{\emph{(a)}} $D$ is reachable from $C$; \textbf{\emph{(b)}} $D(i) = q$ for some $i\in [n]$. Additionally, by a $q$-agent we mean an agent whose state is $q$.

For $q\in Q$, let $f(q)$ denote the minimal $n\in\mathbb{Z}^+$ such that $q$ can occur from $I_n$. If there is no such $n$ at all, set $f(q) = +\infty$. Obviously, $|Q| \ge |f(Q)|$, so it is sufficient to prove that $|f(Q)| \ge \log_2 d + 1$.

Observe that $1 = f(q_{init})$. Hence, $1\in f(Q)$. By definition of 1-awareness, there exists a state $q\in Q_1$ which can occur from $I_d$ (just consider any fair execution starting from $I_d$). On the other hand, no state from $Q_1$ can occur from $I_n$ for $n < d$. Hence, $f(q) = d$ and $d\in f(Q)$. It remains to establish the following lemma.

\begin{lemma}
\label{lemma:gaps}
 Let $a < b$ be two consecutive elements of $f(Q)$. Then $b\le 2a$.
\end{lemma}
Loosely speaking, this lemma asserts that $f(Q)$ does not contain large gaps. Since $1, d\in f(Q)$, it shows that between $1$ and $d$ there must be about $\log_2(d)$ elements of $f(Q)$. In more detail, let $1 = i_1 < i_2 < \ldots < i_k = d$ be elements of $f(Q)$ up to $d$, in the increasing order. By Lemma \ref{lemma:gaps}, we have:
\[i_2 \le 2 i_1, \ldots, i_k \le 2 i_{k - 1}.\]
By taking the product of these inequalities, we obtain:
\[d = i_k \le 2^{k - 1} \cdot i_1 = 2^{k - 1}.\]
Hence, $|f(Q)| \ge k \ge \log_2(d) + 1$.
\begin{proof}[of Lemma \ref{lemma:gaps}]
Consider the minimal $k$ such that some $q\in Q$ with $f(q) = b$ can occur from $I_b$ after $k$ encounters. Note that $k\ge 1$. Indeed, if $k = 0$, then $q = q_{init}$. However, $f(q) = b > a \ge 1$, so $q\neq q_{init}$.

Due to minimality of $k$, a $q$-agent occurs in the last of these $k$ encounters. Consider this agent and also the second agent participating in this encounter. Let their states  prior to the encounter be $q_1$ and $q_2$. We conclude that a $q$-agent can occur whenever we have a $q_1$-agent and a $q_2$-agent in a population (these agents have to be distinct, even when $q_1 = q_2$).

Since $q_1$ and $q_2$ can occur from $I_b$, we have that $f(q_1), f(q_2) \le b$. In turn, since $q_1, q_2$ can occur from $I_b$ in less than $k$ encounters, we have $f(q_1)\neq b$ and $f(q_2) \neq b$, by minimality of $k$. Hence, $f(q_1), f(q_2) \le a$, because  $a$ is the predecessor of $b$ in $f(Q)$. To finish the proof, it is sufficient to show that $f(q) \le f(q_1) + f(q_2)$. In other words, we have to show that $q$ can occur from $I_{f(q_1) + f(q_2)}$. By definition, a $q_1$-agent can occur from  $I_{f(q_1)}$ and a $q_2$-agent can occur from  $I_{f(q_2)}$. Hence, if we have $f(q_1)+ f(q_2)$ agents in the initial state, the first $f(q_1)$ of them are able to produce a $q_1$-agent, while the last $f(q_2)$ of them are able to produce a $q_2$-agent. In turn, these two agents are able to produce a $q$-agent.\qed
\end{proof}

\section{Proof of Theorem \ref{thm:upper_bound}: The First Protocol}
\label{sec:first_protocol}
In this section we establish a 1-aware population protocol with $\log_2(d) + e + O(1)$ states, computing the predicate $R(n) = \mathbb{I}\{n\ge d\}$. Here, $e$ is the number of $1$'s in the binary representation of $d$.

Let $i_1 >  i_2 >\ldots > i_e$ be such that
\[d = 2^{i_1} + 2^{i_2} + \ldots + 2^{i_e}.\]
Imagine that initially every agent holds 1 coin. During the protocol, some agents may run out of coins; we will call these agents \emph{bankrupts}.  At each moment of time, a non-bankrupt agent can hold $1, 2, 4, \ldots, 2^{i_1-1}$ or $2^{i_1}$ coins. Additionally, every bankrupt maintains a counter $k\in\{0, 1, \ldots, e - 1\}$.  Under some circumstances, an agent can come into a special \emph{final} state (informally, this happens when this agent becomes convinced that $n\ge d$). When an agent comes into the final state, it forgets the number of coins it had (this will not be problematic because everything will be decided at this point). So, some agents in the final state might be bankrupt, while the others not. In total, besides the final state, we have  $i_1 + 1 \le \log_2(d) + 1$ states for non-bankrupt agents and $e$ states for bankrupt agents; this is at most $\log_2(d) + e + 2$ states.

We now describe the transitions of our protocol. First, assume that two non-bankrupt agents both having $2^i$ coins meet. If $i = i_1$, then both agents come into the final state. If $i < i_1$, then one of the agents gets all the coins of the other one. That is, one of the agents is left with $2^{i + 1}$ coins, and the other one becomes a bankrupt with $k = 0$. Now, if an agent with $2^i$ coins meets an agent with $2^j$ coins and $j \neq i$, nothing happens. 

Next, we describe transitions that involve bankrupts. If two bankrupts meet, nothing happens. Now, assume that a bankrupt whose counter equals $k$ meets an agent with $2^i$ coins. There are four cases:
\begin{enumerate}
\item if $k < e - 1$ and $i = i_{k + 1}$, then $k$ increments by $1$;
\item if $k = e - 1$ and $i = i_e$, then the bankrupt comes into the final state;
\item if $k > 0$ and $i_k > i > i_{k + 1}$, then the bankrupt comes into the final state;
\item in any other case, the bankrupt sets $k = 0$.
\end{enumerate}
Finally, if an agent is already in the final state, then everybody this agent meets also comes into the final state.

The description of the protocol is finished. To show that this protocol is a 1-aware protocol computing the predicate $R(n) = \mathbb{I}\{n\ge d\}$, it is sufficient to show the following two things:
\begin{itemize}
\item \emph{(soundness)} if $n < d$, then no agent can come into the final state;
\item \emph{(completeness)}  if $n\ge d$, then, after any finite sequence of encounters, it is still possible to bring one of the agents into the final state.
\end{itemize}
Here $n$ is the number of agents in a population.
Indeed, soundness ensures that our protocol satisfies the definition of 1-awareness for $n < d$. Now, consider any $n\ge d$. Take any fair execution $E$ of $n$ agents. We have to show that there exists a moment in $E$, starting from which all agents are always in the final state. Let $C$ be any configuration which occurs infinitely often in $E$. By definition of an execution, $C$ is reachable from $I_n$. Hence, by completeness, there is a configuration $D$ which is reachable from $C$ and which has an agent in the final state. Now, let this agent meet all the other agents. We obtain a configuration $D^\prime$ which is reachable from $D$ and in which all agents are in the final state. By definition of fairness, $D^\prime$ occurs in $E$. Finally, note that once all  agents are in the final state, they will always be in this state.

We start by showing the soundness. Assume for contradiction that there are $n < d$ agents, but one of them came into the final state. First, it could happen if two agents with $2^{i_1}$ coins met. However, the total number of coins is preserved throughout the protocol, and initially there are $n < d = 2^{i_1} + 2^{i_2} + \ldots + 2^{i_e}  < 2 \cdot 2^{i_1}$ coins, contradiction.

Second, it might be that some bankrupt came into the final state. This can happen after an encounter with a non-bankrupt agent. Assume that this non-bankrupt agent held $2^i$ coins. Then there are two options: if $k$ was the value of the counter of our bankrupt agent, then either $k = e - 1$ and $i = i_e$ or $k > 0$ and $i_k > i > i_{k + 1}$. Note that in both cases we have $2^{i_1} + \ldots + 2^{i_k} + 2^i \ge d$. We will show that there must be at least $2^{i_1} + \ldots + 2^{i_k} + 2^i$ distinct coins, and this would be a contradiction.

Consider the counter of our bankrupt. Its current value is $k$. It cannot increase by more than 1 at once. So the last $k$ changes of the counter were as follows: it became equal to 1, then it became equal to 2 and so on, up to a moment when it reached its current value. At the moment when it became equal to 1, our bankrupt saw an agent with $2^{i_1}$ coins. After that, when it became equal to 2, our bankrupt saw an agent with $2^{i_2}$ coins, and so on. In the end, when the counter reached its current value, our bankrupt saw $2^{i_k}$ coins. Additionally, in the very last encounter, it saw $2^i$ coins. We claim these   $2^{i_1} + 2^{i_2} + \ldots + 2^{i_k} + 2^i$ are distinct. To see this, fix any coin. At each moment of time, it belongs to some group of coins. A point it that the size of this group can only increase over time. Now, recall that $2^{i_1} > 2^{i_2} > \ldots > 2^{i_k} > 2^i$. Since our bankrupt first saw a group of $2^{i_1}$ coins, then a smaller group of $2^{i_2}$ coins and so on, none of these coins were seen twice. The soundness is proved.

We now show the completeness. Assume that there are $n\ge d$ agents. Consider any configuration $C$ which is reachable from the initial one. Let $D$ be a configuration which is reachable from $C$ and has the least number of non-bankrupt agents. If in $D$ there are two agents that both have $2^{i_1}$ coins, then we can bring them into the final state. Assume from now on that in $D$ there is at most one agent with $2^{i_1}$ coins. Then no two non-bankrupt agents have the same number of coins in $D$ -- otherwise, one could decrease the number of non-bankrupt agents.

Assume that in $D$ there are $t$ non-bankrupt agents, the first one with $2^{j_1}$ coins, the second one with $2^{j_2}$ coins, and so on.  Here $0 \le j_1, \ldots, j_t \le i_1$. W.l.o.g.~$j_1 > j_2 > \ldots > j_t$. Note that
\[n = 2^{j_1} + 2^{j_2} + \ldots + 2^{j_t} \ge d = 2^{i_1} + 2^{i_2} + \ldots + 2^{i_e}.\]
In particular, $i_1 = j_1$. Moreover, either $t = e$ and $j_1 = i_1, \ldots j_e = i_e$, or there exists $1 \le k < e$ such that $j_1 = i_1, \ldots j_k = i_k$ and $i_k > j_{k+1} > i_{k +1}$.

Now, take any bankrupt (there will be at least one bankrupt already after the first transition). If its counter is not $0$, we reset it to $0$ by pairing our bankrupt with the agent holding $2^{j_1} = 2^{i_1}$ coins. It is now easy to bring this bankrupt into the final state. Indeed, if $t = e$ and $j_1 = i_1, \ldots j_e = i_e$, pair our bankrupt with the agent holding $2^{i_1}$ coins, then with the agent holding $2^{i_2}$ coins, and so on, up to the agent holding $2^{i_e}$ coins. Now, if there exists  $1 \le k < e$, such that $j_1 = i_1, \ldots j_k = i_k$ and $i_k > j_{k+1} > i_{k +1}$, pair our bankrupt with the agent holding $2^{j_1} = 2^{i_1}$ coins, then with the agent holding $2^{j_2} = 2^{i_2}$ coins, and so on, up to the agent holding $2^{j_{k + 1}}$ coins.

\section{Proof of Theorem \ref{thm:upper_bound}: The Second Protocol}
\label{sec:second_protocol}

In this section we establish a 1-aware population protocol with $\log_2(d) + z + O(1)$ states computing the predicate $R(n) = \mathbb{I}\{n\ge d\}$. Here, $z$ is the number of $0$'s in the binary representation of $d - 1$.

As a warm-up, we first consider  $d = 2^{k + 1} - 1$. In this case, $z = 1$. The protocol from the previous section requires about $2\log_2 d$ states for such $d$. We present a simple protocol which only needs $\log_2 d + O(1)$ states for such $d$.

\subsection{Warm-up: case $d = 2^{k + 1} - 1$.}

Again,  initially each agent holds 1 coin. As before, we distinguish between bankrupt and non-bankrupt agents. A non-bankrupt agent can hold $1, 2, \ldots, 2^{k - 1}$ or $2^k$ coins. Thus, there are $k + 1$ possible states of non-bankrupt agents, 1 state indicating bankrupts, and also 1 final state -- in total, $k + 3 = \log_2 d+ O(1)$ states.

We now describe the transitions of the protocol. Assume that two agents both having $2^i$ coins for some $0 \le i < k - 1$ meet. Then, as in the previous section, one of them gets $2^{i + 1}$ coins and the other one becomes bankrupt. Now, when two agents both having $2^{k - 1}$ coins meet, one of them gets $2^k$ coins and the other one gets 1 coin ``out of nowhere''. When two agents with $2^k$ coins meet, both of them come into the final state. Finally, if an agent is already in the final state, then everybody this agent meets also comes into the final state. All the other encounters do not change states of agents.

The rest of the argument has the same two parts -- ``soundness'' and ``completeness''.  ``Soundness'' means that if $n$, the total number of agents, is smaller than $d$, then no agent can come into the final state. ``Completeness'' means that if $n\ge d$, then, after any sequence of encounters, it is still possible to bring one of the agents into the final state. Similarly to the argument from the previous section,  ``soundness'' and  ``completeness'' imply that our protocol is a 1-aware protocol computing $R(n) = \mathbb{I}\{n\ge d\}$.

Let us start with the soundness. Assume that $n < d$. We claim that  an ``out of nowhere'' coin may occur at most once. Indeed, consider the first time it occurs. At this moment, one of the agents gets $2^k$ coins. Nothing happens with these $2^k$ coins unless we get one more agent with $2^k$ coins. However, all the other agents in total have $(n - 2^k) + 1 < (2^{k + 1} - 1 - 2^k) + 1 = 2^k$ coins. Thus, from now on it is impossible to get two agents with $2^{k - 1}$ coins. In particular,  it is impossible to get a coin ``out of nowhere''. This means that the total number of coins never exceeds $n + 1 < 2^{k + 1}$. However, to bring somebody into the final state, we must have at least $2^{k + 1}$ coins. The soundness is proved.

Let us now show the completeness. Assume that $n\ge d$. Let $C$ be any configuration, reachable from the initial configuration of $n$ agents. Assume for contradiction that no configuration with an agent in the final state is reachable from $C$. Let $D$ be a configuration which is reachable from $C$ and has the most coins in total (as any agent can hold up to $2^k$ coins, the total number of coins is bounded by $n2^k$). Next, let $D^\prime$ be a configuration which is reachable from $D$ and has the least number of non-bankrupts. We have at most 1 agent with $2^k$ coins in $D^\prime$ -- otherwise we could reach the final state. Also, in $D^\prime$ there is at most one agent with $2^{k - 1}$ coins -- otherwise we could increase the total number of coins. Similarly, for every $0 \le i < k -1 $, there is at most one agent with $2^i$ coins -- otherwise one could decrease the number of non-bankrupts. Thus, we have at most $1 + 2 + \ldots + 2^k = d$ coins in total. Initially, there are $n \ge d$ coins. The total number of coins does not decrease in our protocol. Hence, in $D^\prime$ there must be exactly $d = 1 + 2 + \ldots + 2^k$ coins. In particular, in $D^\prime$ there must be an agent with $2^k$ coins. When  an agent with $2^k$ coins occurs, we also get a coin ``out of nowhere''. This means that in $D^\prime$ the total number of coins is bigger than in the initial configuration. That is, initially there were at most $d - 1$ coins, contradiction.

\subsection{General case}

We assume that $d$ is not a power of $2$ (otherwise we could use the protocol from Section \ref{sec:first_protocol}). 
Let $2^k$ be the largest power of $2$ below $d$. Define $a = 2^{k + 1} - d$. Observe that:
\[2^{k + 1} - 1 = \underbrace{11\ldots1}_{k +1} = (d - 1) + a\]
Since $2^k < d < 2^{k + 1}$, there are $k + 1$ digits in the binary representation of $d - 1$. Hence, the number of $1$'s in the binary representation of $a$ equals the number of $0$'s in the binary representation of $d - 1$ (that is, equals $z$).

Assume that
\begin{equation}
\label{a}
a = 2^{b_1} + 2^{b_2} + \ldots + 2^{b_z},
\end{equation}
where $b_1 > b_2 > \ldots > b_z$. Note that $a = 2^{k + 1} - d < 2^{k + 1} - 2^k = 2^k$. Hence, $b_1 < k$.

The protocol in the general is essentially the same as for the case $d = 2^{k + 1} - 1$, except that instead of just 1 coin ``out of nowhere'' we get $a$ coins ``out of nowhere'' every time two agents with $2^{k-1}$ coins meet. However, there will be additional technical difficulties, as $a$ might not be a power of 2. 

In more detail, a non-bankrupt agent may have 
\[s\in\{1, 2, \ldots, 2^k, 2^{b_1} + 2^{b_2}, 2^{b_1} + 2^{b_2} + 2^{b_3}, \ldots, 2^{b_1} + 2^{b_2} + \ldots + 2^{b_z} = a\} \mbox{ coins}.\]
Thus, a non-bankrupt agent can be in one of the $k + z$ states. Taking into the account the state indicating bankrupts and the final state, in total we have $k + z + 2 \le \log_2 d + z + O(1)$ states. We will call agents that hold $ 2^{b_1} + 2^{b_2} + \ldots + 2^{b_i}$ coins for some $i > 1$ \emph{non-standard}.

Let us now describe transitions of the protocol.
When two agents with $2^i$ coins meet, where $0 \le i < k - 1$, one of them gets $2^{i + 1}$ coins and the other one becomes bankrupt. When two agents with $2^{k - 1}$ coins meet, one of them gets $2^k$ coins and the other one gets $a$ coins ``out of nowhere''. When two agents with $2^k$ coins meet, both of them come into the final state. Now, when a non-standard agent with $2^{b_1} + \ldots + 2^{b_i}$ meets a bankrupt, this bankrupt gets $2^{b_i}$ coins, and the non-standard agent is left with $2^{b_1} + \ldots + 2^{b_{i - 1}}$ coins (if $i = 2$, the non-standard agent becomes standard). Finally, if an agent is already in the final state, then everybody this agent meets also comes into the final state. All the other encounters do not change the states of agents.

Let us now show the soundness of our protocol. Assume that the number of agents is $n < d$. We show that no agent can be brought into the final state. For that, we first show that we can get $a$ coins out of nowhere at most once. Indeed, consider the first time this happened. We get one agent with $2^k$ coins. Other agents have $n - 2^k + a = n - 2^k + (2^{k+1} - d) < 2^k$ coins in total (the inequality holds because $n < d$). Thus, we will never have two agents with $2^{k - 1}$ coins again. Thus, in any execution, the total number of coins never exceeds $n + a = n + (2^{k+1} - d) < 2^{k + 1}$. However, to bring somebody into the final state, we 
must have two agents with $2^k$ coins.

Let us now show the completeness of our protocol. Assume that $n\ge d$. Let $C$ be any configuration, reachable from the initial one. Assume for contradiction that no configuration with an agent in the final state is reachable from $C$. Let $C_1$ be a configuration which is reachable from $C$ and has the most coins in total (as before, this number is bounded by $n2^k$). Next, let $C_2$ be a configuration which is reachable from $C_1$ and minimizes the following parameter:
\begin{align*}
p &= \mbox{the number of standard non-bankrupt agents}\\
&+3\times\mbox{the number of coins belonging to non-standard agents}.
\end{align*}
In $C_2$ there is at most one agent with $2^k$ coins -- otherwise, we could reach the final state. There is also at most one agent with $2^{k - 1}$ coins -- otherwise, one could increase the total number of coins. Similarly, for any $0 \le i < k - 1$, there is at most one agent in $C_2$ with $2^i$ coins (otherwise, by pairing two agents with $2^i$ coins, one could decrease $p$).

If there is no agent with $2^k$ coins in $C_2$, then we never got $a$ coins out of nowhere on our path to $C_2$. Indeed, when we get $a$ coins out of nowhere, we get an agent with $2^k$ coins, and nothing happens with this agent unless the final state is reached. So there are 0 non-standard agents in $C_2$ (they are created only when we get coins out of nowhere). Hence, there are at most $1 + 2 + \ldots + 2^{k - 1} < 2^k < d$ coins in $C_2$, contradiction. 

Hence, in $C_2$ there is exactly one agent with $2^k$ coins. Clearly, this also means that on our path to $C_2$ we got $a$ coins out of nowhere exactly once. This is because the only transition creating an agent with $2^k$ coins is the transition creating $a$ coins out of nowhere. Indeed, other transitions with standard agents create smaller power of 2, and transitions with non-standard agents involve at most $a < 2^k$ coins. We conclude that, first, in $C_2$ there is exactly one agent with $2^k$ coins, second, there are $n + a$ coins in total, and third, there is at most 1 non-standard agent (it could be created only once, when we got $a$ coins out of nowhere).

Assume first that all agents in $C_2$ are standard. Then $n + a \le 1 + 2 + \ldots + 2^k = 2^{k + 1} - 1$. Hence, $n \le 2^{k + 1} - 1 - a = d - 1$, contradiction.

Now, assume that in $C_2$ there is exactly one non-standard agent who holds $2^{b_1} + \ldots + 2^{b_i}$ coins, $i > 1$. Let us show that in $C_2$ there exists a bankrupt agent. Indeed, assume for contradiction that all agents in $C_2$ are non-bankrupt. Now, leave every agent with exactly one coin. There will be exactly $n$ coins. That is, exactly $a$ coins were taken. However, from the agent with $2^k$ coins we took $2^k - 1$ coins. Additionally, we took at least 1 coin from the non-standard agent. Hence, we took at least $2^k$ coins. Since $a < 2^k$, we obtain a contradiction.

Now, pair the non-standard agent with any bankrupt agent. We claim that the parameter $p$ will decrease (this will be a contradiction with the definition of $C_2$). Indeed, as a result, we get at most 2 new standard non-bankrupt agents. However, the number of coins belonging to the non-standard agent decreases by at least 1. Therefore, $p$ decreases.

\subsubsection*{Acknowledgements} I would like to thank Karoliina Lehtinen and K.~S.~Thejaswini for discussions on population protocols.

%
%
%
%
%
%

\begin{thebibliography}{10}

\bibitem{alistarh2017time}
{\sc Alistarh, D., Aspnes, J., Eisenstat, D., Gelashvili, R., and Rivest,
  R.~L.}
\newblock Time-space trade-offs in population protocols.
\newblock In {\em Proceedings of the twenty-eighth annual ACM-SIAM symposium on
  discrete algorithms\/} (2017), SIAM, pp.~2560--2579.

\bibitem{angluin2006computation}
{\sc Angluin, D., Aspnes, J., Diamadi, Z., Fischer, M.~J., and Peralta, R.}
\newblock Computation in networks of passively mobile finite-state sensors.
\newblock {\em Distributed computing 18}, 4 (2006), 235--253.

\bibitem{angluin2008fast}
{\sc Angluin, D., Aspnes, J., and Eisenstat, D.}
\newblock Fast computation by population protocols with a leader.
\newblock {\em Distributed Computing 21}, 3 (2008), 183--199.

\bibitem{angluin2007computational}
{\sc Angluin, D., Aspnes, J., Eisenstat, D., and Ruppert, E.}
\newblock The computational power of population protocols.
\newblock {\em Distributed Computing 20}, 4 (2007), 279--304.

\bibitem{aspnes2009introduction}
{\sc Aspnes, J., and Ruppert, E.}
\newblock An introduction to population protocols.
\newblock {\em Middleware for Network Eccentric and Mobile Applications\/}
  (2009), 97--120.

\bibitem{Blondin2020SuccinctPP}
{\sc Blondin, M., Esparza, J., Genest, B., Helfrich, M., and Jaax, S.}
\newblock Succinct population protocols for presburger arithmetic.
\newblock In {\em STACS\/} (2020).

\bibitem{blondin_et_al:LIPIcs:2018:8511}
{\sc Blondin, M., Esparza, J., and Jaax, S.}
\newblock {Large Flocks of Small Birds: on the Minimal Size of Population
  Protocols}.
\newblock In {\em 35th Symposium on Theoretical Aspects of Computer Science
  (STACS 2018)\/} (Dagstuhl, Germany, 2018), R.~Niedermeier and B.~Vall{\'e}e,
  Eds., vol.~96 of {\em Leibniz International Proceedings in Informatics
  (LIPIcs)}, Schloss Dagstuhl--Leibniz-Zentrum fuer Informatik,
  pp.~16:1--16:14.

\bibitem{10.1145/3465084.3467912}
{\sc Czerner, P., and Esparza, J.}
\newblock Lower bounds on the state complexity of population protocols.
\newblock In {\em Proceedings of the 2021 ACM Symposium on Principles of
  Distributed Computing\/} (New York, NY, USA, 2021), PODC'21, Association for
  Computing Machinery, p.~45–54.

\bibitem{czerner_et_al:LIPIcs.SAND.2022.11}
{\sc Czerner, P., Guttenberg, R., Helfrich, M., and Esparza, J.}
\newblock {Fast and Succinct Population Protocols for Presburger Arithmetic}.
\newblock In {\em 1st Symposium on Algorithmic Foundations of Dynamic Networks
  (SAND 2022)\/} (Dagstuhl, Germany, 2022), J.~Aspnes and O.~Michail, Eds.,
  vol.~221 of {\em Leibniz International Proceedings in Informatics (LIPIcs)},
  Schloss Dagstuhl -- Leibniz-Zentrum f{\"u}r Informatik, pp.~11:1--11:17.

\bibitem{diamadi2001simple}
{\sc Diamadi, Z., and Fischer, M.~J.}
\newblock A simple game for the study of trust in distributed systems.
\newblock {\em Wuhan University Journal of Natural Sciences 6}, 1 (2001),
  72--82.

\bibitem{gillespie1977exact}
{\sc Gillespie, D.~T.}
\newblock Exact stochastic simulation of coupled chemical reactions.
\newblock {\em The journal of physical chemistry 81}, 25 (1977), 2340--2361.

\end{thebibliography}
\end{document}